Corresponding Author: Professor Maria Vallet-Regí,

Corresponding Author's Institution: Universidad Complutense

First Author: Marina Martínez-Carmona

Order of Authors: Marina Martínez-Carmona; Daniel Lozano; Montserrat Colilla; Maria Vallet-Regí

Abstract: A novel multifunctional nanodevice based in doxorubicin (DOX)-loaded mesoporous silica nanoparticles (MSNs) as nanoplatforms for the assembly of different building blocks has been developed for bone cancer treatment. These building blocks consists of: i) a polyacrylic acid (PAA) capping layer grafted to MSNs via an acid-cleavable acetal linker, to minimize premature cargo release and provide the nanosystem of pH-responsive drug delivery ability; and ii) a targeting ligand, the plant lectin concanavalin A (ConA), able to selectively recognize, bind and internalize owing to certain cell-surface glycans, such as sialic acids (SA), overexpressed in given tumor cells. This multifunctional nanosystem exhibits a noticeable higher internalization degree into human osteosarcoma cells (HOS), overexpressing SA, compared to healthy preosteoblast cells (MC3T3-E1). Moreover, the results indicate that small DOX loading (2.5 µg mL-1) leads to almost 100% of osteosarcoma cell death in comparison with healthy bone cells, which significantly preserve their viability. Besides, this nanodevice has a cytotoxicity on tumor cells 8-fold higher than that caused by the free drug. These findings demonstrate that the synergistic combination of different building blocks into a unique nanoplatform increases antitumor effectiveness and decreases toxicity towards normal cells. This line of attack opens up new insights in targeted bone cancer therapy.

Manuscript submitted to

Acta Biomaterialia

**New Paradigm in Targeted Bone Cancer Therapy**

Marina Martínez-Carmona,[a,b] Daniel Lozano,[a,b] Montserrat Colilla,[a,b*]

María Vallet-Regí[a,b*]

[a] Dpto. Química Inorgánica y Bioinorgáni*ca* Universidad Complutense de Madrid. Instituto de Investigación Sanitaria Hospital 12 de Octubre i+12. Plaza Ramón y Cajal s/n, 28040 Madrid, Spain.

[b] CIBER de Bioingeniería, Biomateriales y Nanomedicina, CIBER-BBN, Madrid, Spain.

[*] Corresponding authors: Fax: +34 394 1786; Tel.: +34 91 394 1843; E-mail addresses: vallet@ucm.es (M. Vallet-Regí) and mcolilla@ucm.es (M. Colilla)

**Abstract**

A novel multifunctional nanodevice based in doxorubicin (DOX)-loaded mesoporous silica nanoparticles (MSNs) as nanoplatforms for the assembly of different building blocks has been developed for bone cancer treatment. These building blocks consists of: i) a polyacrylic acid (PAA) capping layer grafted to MSNs *via* an acid-cleavable acetal linker, to minimize premature cargo release and provide the nanosystem of pH-responsive drug delivery ability; and ii) a targeting ligand, the plant lectin concanavalin A (ConA), able to selectively recognize, bind and internalize owing to certain cell-surface glycans, such as sialic acids (SA), overexpressed in given tumor cells. This multifunctional nanosystem exhibits a noticeable higher internalization degree into human osteosarcoma cells (HOS), overexpressing SA, compared to healthy preosteoblast cells (MC3T3-E1). Moreover, the results indicate that small DOX loading (2.5 µg mL$^{-1}$) leads to almost 100% of osteosarcoma cell death in comparison with healthy bone cells, which significantly preserve their viability. Besides, this nanodevice has a cytotoxicity on tumor cells 8-fold higher than that caused by the free drug. These findings demonstrate that the synergistic combination of different building blocks into a unique nanoplatform increases antitumor effectiveness and decreases toxicity towards normal cells. This line of attack opens up new insights in targeted bone cancer therapy.

**1. Introduction**

The major constraint of conventional chemotherapy for cancer treatment is the lack of specificity of cytotoxic drugs. This constitutes a serious threat to healthy tissues, resulting in a significant decrease of the efficacy and provoking the apparition of side-effects in the patient associated to systemic toxicity[1]. The emergence of nanotechnology has transformed this scenario owing to the development of nanocarriers for therapeutic agents[2]. Particularly, in the last two decades much research effort has been devoted to build nanocarriers for tumor-targeted stimuli-responsive drug delivery[3-7].

Among nanocarriers, mesoporous silica nanoparticles (MSNs) are receiving growing interest since they exhibit unique properties such as large surface areas and pore volumes, which provide high loading capability, customized sizes, morphology and pore diameters, robustness and easy functionalization[8-19]. Moreover, they exhibit low cytotoxicity[20] and good hemocompatibility[21]. These features provide exceptional opportunities to host different therapeutic payloads. Hence, the tunable surface of MSNs chemistry allows the attachment of organic functionalities such as pore blocking agents to avoid premature cargo release, or targeting ligands to guide MSNs towards diseased tissues.

In this sense, during the last few years, a wide range of stimuli-responsive drug delivery systems, achieving release profiles with spatial, temporal and dosage control have been reported[4,15,17]. The performance of these smart nanosystems relies on triggering drug release at the target site by using moieties that are sensitive to external stimuli (light, magnetic fields, electric fields, ultrasounds, etc.)[22-25], internal stimuli (variations in pH, redox potential, or the concentrations of enzymes or specific analytes)[26,27], or combination of both[28]. However, the side effects in most of these nanosystems cannot be disregarded in practice because these particles frequently lack of specific cancer cell targeting capability and can be also internalized by normal cells. In the last decade, considerable research effort has been committed to develop tumor-targeted stimuli-responsive drug delivery systems. Although diverse targeting ligands, such as transferrin[23,29], certain peptides[30], hyaluronic acid[31,32], or folic acid[26,33], have been incorporated into stimuli-responsive MSNs, the development of highly selective and efficient tumor-targeted smart drug delivery nanodevices remains a tremendous challenge.

Herein we have designed and developed a multifunctional nanodevice based in doxorubicin (DOX)-loaded MSNs acting as nanoplatforms for the assembly of different building blocks: i) a pH-sensitive layer consisting in polyacrylic acid (PAA) anchored to the MSNs surface *via* an acid-cleavable acetal linker[34]. PAA was used as pore blocking agent due to its good properties such as biocompatibility, hydrophilicity and abundance of carboxylic groups prone to experience easy functionalization[35]; and ii) a targeting ligand, the lectin concanavalin A (ConA) grafted to PAA, to increase the selectivity of the nanocarrier towards cancer cells and preserve the viability of healthy cells. ConA was chosen as targeting ligand owing the capability of this plant lectin to selectively recognize and bind to cell-surface glycans, which are frequently overexpressed in cancer cells[36,37].

The pH-responsive drug delivery performance of the nanosystem was evaluated in vial using $[Ru(bipy)_3]^{2+}$ as model molecule. The results demonstrate that in acidic conditions (pH 5.3) mimicking those of endo/lysosomes drug released is faster than in physiological conditions (pH 7.4). Then, the capacity of this multifunctional nanodevice to selectively eradicate human osteosarcoma cancer cells (HOS), overexpressing sialic acid (SA) glycans as ConA receptors, compared to healthy preosteoblastic cells (MC3T3-E1), non-overexpressing SA, was in vitro evaluated. The results indicate the preferential internalization of the nanosystems into tumor cells and that only very small DOX concentrations are needed to almost completely eradicate osteosarcoma cells meanwhile preserving the viability of healthy bone cells. This resulted in a 100% killing efficacy against tumor cells in comparison with normal cells, as well as 8-fold higher tumor cytotoxicity than that caused by free DOX. We envision this novel multifunctional nanodevice as a potential candidate to be incorporated into a nanomedicines library for targeted bone cancer therapy.

## 2. Materials and Methods

### 2.1 Reagents

Tetraethylorthosilicate (TEOS, 98%), n-cetyltrimethylammo-nium bromide (CTAB, ≥ 99%), sodium hydroxide (NaOH, ≥ 98%), ammonium nitrate ($NH_4NO_3$, ≥ 98%), sodium carbonate ($Na_2CO_3$, ≥ 99,5%), hydrochloric acid (HCl, 37%), fluorescein 5(6)-isothiocyanate (FITC, ≥ 98%), (3-aminopropyl) triethoxysilane (APTES, ≥ 98%), N,N'-dicyclohexylcarbodiimide (DCC, 99%), N-hydroxysuccinimide (NHS, 98%), poly(acrylic acid) partial sodium salt solution (PAA, average Mw ~ 240,000 by GPC, 25 wt.% in $H_2O$), phosphate-buffered saline (PBS, 10x), phosphotungstic acid hydrate (PTA, reagent grade), Concanavalin A from Canavalia ensiformis (Jack bean) (ConA,

Type VI lyophilized powder), tris(bipyridine)ruthenium(II) chloride ([Ru(bipy)$_3$]Cl$_2$), and doxorubicin hydrochloride (DOX, European Pharmacopoeia) were purchased from Sigma-Aldrich (St. Louis, USA). 3-(Triethoxysiyl)propylsuccinic anhydride (TEPSA, 95%) was purchased from ABCR (Karlsruhe, Germany) and 3,9-Bis(3-aminopropyl)-2,4,8,10-tetraoxaspiro [5.5] undecane (ATU) was purchased from TCY (Tokyo, Japan). All other chemicals were purchased from Panreac Química SLU (Castellar del Valles, Barcelona, Spain) inc: absolute ethanol, acetone, dimethyl sulfoxide (DMSO), etc. All reagents were used as received without further purification. Ultrapure deionized water with resistivity of 18.2 MΩ was obtained using a Millipore Milli-Q plus system (Millipore S.A.S, Molsheim, France).

**2.2 Characterization techniques**

Powder X-Ray Diffraction (XRD) experiments were performed in a Philips X'Pert diffractometer equipped with Cu Kα radiation (wavelength 1.5406 Å) (Philips Electronics NV, Eindhoven, Netherlands). XRD patterns were collected in the 2θ range between 0.6° and 8° with a step size of 0.02° and counting time of 5 s per step. Thermogravimetric (TG measurements were performed in a Perkin Elmer Pyris Diamond TG/DTA (California, USA), with 5 ºC min$^{-1}$ heating ramps, from room temperature to 600 ºC. Fourier transform infrared spectroscopy (FTIR) was carried out in a Nicolet (Thermo Fisher Scientific, Waltham, MA, USA) Nexus spectrometer equipped with a Goldengate attenuated total reflectance (ATR) accessory (Thermo Electron Scientific Instruments LLC, Madison, WI USA). Morphology, mesoestructural order and nanoparticles functionalization were studied by High Resolution Transmission Electron Microscopy (HRTEM) with a JEOL JEM 3000F instrument operating at 300 kV, equipped with a CCD camera (JEOL Ltd., Tokyo, Japan). Sample preparation was performed by dispersing in distilled water and subsequent deposition onto carbon-coated copper grids. A 1% PTA solution (pH 7.0) was used as staining agent in order to visualize the organic coating around MSNs.

To determine the evolution of the size and surface charge of nanoparticles by dynamic light scattering (DLS) and zeta (ζ-potential measurements, respectively, a Zetasizer Nano ZS (Malvern Instruments, United Kingdom) equipped with a 633 nm "red" laser was used. DLS measurements were directly recorded in ethanolic colloidal suspensions. ζ-potential measurements were recorded in aqueous colloidal suspensions. For this purpose 1 mg of nanoparticles was added to 10 mL of solvent followed by 5 min of sonication to obtain a homogeneous suspension. In both cases measurements were recorded by placing 1 mL of suspension (0.1 mg mL$^{-1}$) in DTS1070 disposable folded capillary cells (Malvern Instruments). The textural properties of the materials were determined by $N_2$ adsorption porosimetry by using a Micromeritics ASAP 2020 (Micromeritics Co., Norcross, USA). To perform the $N_2$ measurements, 20-30 mg of each sample was previously degassed under vacuum for 24 h at 40 ºC temperature. The surface area ($S_{BET}$) was determined using the Brunauer-Emmett-Teller (BET) method and the pore volume ($V_P$) was estimated from the amount of $N_2$ adsorbed at a relative pressure around 0.97. The pore size distribution between 0.5 and 40 nm was calculated from the adsorption branch of the isotherm by means of the Barrett-Joyner-Halenda (BJH) method. The mesopore size ($D_P$) was determined from the maximum of the pore size distribution curve. To evaluate the different carbon environments, $^1H \rightarrow {}^{13}C$ CP (cross-polarization)/ MAS (magic angle spinning) solid-state nuclear magnetic resonance (NMR) measurements were performed in a Bruker AV-400-WB spectrometer (Karlsruhe, Germany) operating at 75.45 MHz. Solid samples were placed in a 4 mm zirconia rotor and spun at 12 kHz. Chemical shifts (δ) of 13C were externally referred to glycine at δ = 0.0 ppm. Time periods between successive accumulations were 3 ms and *ca* 15,000 scans were collected.

**2.3 Synthesis of pure-silica MSNs (MSN)**

Bare MSNs, denoted as MSN, were synthesized by the modified Stöber method using TEOS as silica source in the presence of CTAB as structure directing agent. Briefly, 1 g of CTAB, 480 mL of $H_2O$ and 3.5 mL of NaOH (2 M) were added to a 1,000 mL round-bottom flask. The mixture was heated to 80 ºC and magnetically stirred at 600 rpm. When the reaction mixture was stabilized at 80 ºC, 5 mL of TEOS were added dropwise at 0.33 mL min$^{-1}$ rate. The white suspension obtained was stirred during further 2 h at 80 ºC. The reaction mixture was centrifuged and washed three times with water and ethanol. Finally the product was dried under vacuum at 40 ºC. The surfactant was removed by ionic exchange by soaking 1 g of nanoparticles in 500 mL of a $NH_4NO_3$ solution (10 mg mL$^{-1}$) in ethanol (95%) at 65 ºC overnight under magnetic stirring. The nanoparticles were collected by centrifugation, washed twice with water and twice with ethanol and dried under vacuum at 40 ºC.

For cellular internalization studies fluorescein-labeled MSN were synthesized. For this purpose, 1 mg FITC and 2.2 μL APTES were dissolved in 100 μL ethanol and left reacting for 2 h. Then the reaction mixture was added with the 5 mL of TEOS as previously described.

### 2.4. Functionalization of MSN with carboxylic acid groups (MSN$_{COOH}$)

With the aim of selectively grafting carboxylic acid groups to the external surface of MSN while keeping the mesopore intact for cargo loading, functionalization was carried out in MSN before being submitted to the surfactant extraction process. Thus, 500 mg of CTAB-containing MSN were placed in a three-neck round bottom flask and dried at 80 ºC under vacuum for 24 h. Then, 125 mL of dry toluene were added and the flask was placed in an ultrasonic bath to favour the suspension of nanoparticles. After that 300 μL of TEPSA were added, keeping the reaction under nitrogen atmosphere at 90 ºC for 24 h. Next, 40 mL of slightly acidified water were added in order to hydrolyze the succinic groups to carboxylic acid groups[38]. The nanoparticles were collected by centrifugation, washed three times with ethanol and dried under vacuum at 40 ºC. For deep characterization of this carboxylic acid MSNs, the surfactant removal was accomplished by solvent extraction as above mentioned, affording MSN$_{COOH}$.

### 2.5. Functionalization of MSN with a pH-cleavable linker (MSN$_{ATU}$)

The grafting of ATU to the outermost surface of surfactant-containing MSN$_{COOH}$ was attained by adding 300 mg of these MSNs to a solution of EDC (480 mg) and NHS (180 mg) in 50 mL PBS 1x. After 30 min of magnetic stirring 2 g of ATU were added to the suspension and the mixture was reacted overnight. The product was filtered, washed with water and the surfactant removal was then carried out, affording MSN$_{ATU}$. This sample was left to dry under vacuum at 40 ºC.

**2.6. Capping of MSN with PAA (MSN$_{PAA}$)**

The first step was to investigate the amount of acid-sensitive polymer that led to the most effective and uniform coating while avoiding nanoparticles aggregation. With this goal in mind 10 mg of MSN$_{ATU}$ was suspended in 1 mL PBS (10 mM). Then 0.36, 3.6 or 36 µL PAA were added to the suspension and the resulting PAA-coated MSNs were characterized by TEM and their molecule release performance was preliminary evaluated in vial, as it will be described below. The most effective and most uniform coating (Fig. S1, Supporting Information) and the best preliminary in vial release behavior (results not shown) were achieved for the lowest polymer amount, i.e. 0.36 µL. Thus, this amount was chosen to synthetize PAA-coated nanoparticles, named MSN$_{PAA}$.

**2.7. ConA grafting to MSN$_{PAA}$ (MSN$_{ConA}$)**

16 mg of MSN$_{PAA}$ were placed in a vial and suspended in 2 mL of PBS pH 7.4. After that, 30 mg of EDC were added and the mixture was stirred at R.T. for 40 min. Then 14 mg of NHS were added and the reaction was stirred for 10 min before adding 30 mg of ConA and left to react overnight at R.T. Finally, samples were filtered, washed twice with PBS pH 7.4 and dried under vacuum at 25 ºC.

**2.8. Cargo loading**

40 mg of MSN$_{ATU}$ were placed in a dark glass vial and dried at 80 ºC overnight under vacuum. Then, 6 mL of [Ru(bipy)$_3$]Cl$_2$ (a model molecule) or DOX (an antitumor drug) aqueous solution (10 or 3 mg mL$^{-1}$ respectively) were added and the suspension was stirred at R.T. for 24 h. After that, 2.15 µL of PAA were added and stirred for 10 min before adding 96 mg of EDC and 36 mg of NHS and finally allowed to react overnight at R.T. Next, samples were filtered and washed twice with PBS pH 7.4 in order to remove the [Ru(bipy)$_3$]Cl$_2$ or DOX adsorbed on the external surface of the nanoparticles. Finally, the products were dried under vacuum at 25 ºC. The grafting of ConA was then carried out by following the procedure previously described, leading to MSN$_{ConA}$@[Ru(bipy)$_3$]$^{2+}$ or MSN$_{ConA}$@DOX nanoparticles.

The quantity of [Ru(bipy)$_3$]$^{2+}$ loaded into the nanosystem was estimated from the maximum amount of dye released (*ca* 100%) after 24 h of delivery assay. Thus, the amount of dye released at each time point was normalized to such amount. On the other hand, the amount of DOX loaded in nanoparticles was determined from the difference between the fluorescence measurements of the initial and the recovered filtrate solutions in the step previous to ConA grafting. In both cases, two different calibration lines (one at each tested pH) were performed to eliminate the contribution of nonspecific release by pH effect.

**2.9. In vial cargo release assays**

To investigate the pH-responsive drug release performance of $MSN_{ConA}$, $[Ru(bipy)_3]^{2+}$ was chosen as model molecule and in vial time-based fluorescence release experiments at two different pH values, i.e. 7.4 and 5.3, were carried out. Fluorescence measurements were recorded on a PTI QuantaMaster 400 system featuring a JYF-FLUOROMAX-4 compact spectrofluorimeter single grating excitation and emission monochromator with a photomultiplier detector (PMT R928P) and an automated four-position thermostated cuvette-holder (FL-1011) (PTI, Photon Technology International, HORIBA Jobin Yvon GmbH, Germany). For temperature settings, a MTB-IFI-156-5251 refrigerated bath circulator with a peltier was used. $[Ru(bipy)_3]^{2+}$ was excited with 451 nm and showed a maximum of emission at 619 nm (excitation slit 0.38 mm, emission slit 0.38 mm, integration 0.5 sec). For release experiments, the procedure reported by Bein and co-workers was used[39]. Thus, 170 μL of a 2 mg mL$^{-1}$ nanoparticles suspension was filled into a reservoir cap sealed with a dialysis membrane (molecular weight cut-off 12000 g mol$^{-1}$), allowing released dye molecules to pass into the fluorescence cuvette (which was completely filled with PBS at a given pH) while the relatively large particles are held back. Experiments were performed at a temperature of 37 °C.

**2.10. Cell cultures**

Cell culture tests were performed using the well-characterized mouse preosteoblastic cell line MC3T3-E1 (subclone 4, CRL-2593; ATCC, Mannassas, VA) and HOS cells derived from a human osteosarcoma (CRL-1543; ATCC, Mannassas, VA). The tested nanoparticles were placed into each well of 6- or 24-well plates (Corning, CULTEK, Madrid, Spain) after cell seeding. MC3T3-E1 and HOS cells were then plated at a density of 20,000 cells cm$^{-2}$ in 1 mL of α-minimum essential medium (MEM) or Dulbecco's modified Eagle's medium (DMEM, Sigma Chemical Company), respectively, containing 10% of heat-inactivated fetal bovine serum (FBS, Thermo Fisher Scientific) and 1% penicillin–streptomycin (BioWhittaker Europe, Verviers, Belgium) at 37 ºC in a humidified atmosphere of 5% $CO_2$, and incubated for different times. Some wells contained no nanoparticles as controls.

**2.11. Cell viability**

Cell growth was analyzed using the CellTiter 96® AQueous Assay (Promega, Madison, WI, USA), a colorimetric method for determining the number of living cells in culture. Briefly, both type of cells were cultured as described above without (control) or with the tested materials and/or different concentrations of ConA for several times. At each time, 40 μL of CellTiter 96 AQueous One Solution Reagent (containing 3-(4,5-dimethythizol-2-yl)-5-(3-carboxymethoxyphenyl)-2-(4-sul-fophe-nyl)-2H-tetrazoliumsalt (MTS) and an electron coupling reagent (phenazine ethosulfate) that allows its combination with MTS to form a stable solution was added to each well and incubated for 4 h. The absorbance at 490 nm was then measured in a Unicam UV-500 UV–visible spectrophotometer (Thermo Spectronic, Cambridge, UK).

**2.12. Flow cytometry studies**

MC3T3-E1 and HOS cells were cultured in each well of a 6-well plate. After 24h, the cells were incubated at different times in the absence or presence of the tested nanoparticles (100 μg mL$^{-1}$). After 2 h, cells were washed twice with PBS and incubated at 37 ºC with trypsin–EDTA solution (Sigma-Aldrich) for cell detachment. The reaction was stopped with culture medium after 5 min and cells were centrifuged at 1,000 rpm for 10 min and resuspended in fresh medium. Then, the surface fluorescence of the cells was quenched with trypan blue (0.4%) to confirm the presence of an intracellular, and therefore internalized, fluorescent signal. Flow cytometry measurements were performed at an excitation wavelength of 488 nm, green fluorescence was measured at 530 nm (FL1). The trigger was set for the green fluorescence channel (FL1). The conditions for the data acquisition and analysis were established using negative and positive controls with the CellQuest Program of Becton–Dickinson and these conditions were maintained during all the experiments.

Each experiment was carried out three times and single representative experiments are displayed. For statistical significance, at least 10,000 cells were analyzed in each sample in a FACScan machine (Becton, Dickinson and Company, USA) and the mean of the fluorescence emitted by these single cells was used.

**2.13. Fluorescence microscopy**

Cells were incubated with the MSNs (100 μg mL$^{-1}$) for 2 h. Each well was washed with cold PBS for three more times to get rid of the nanoparticles no internalized into the cells, and then fixed with 75% ethanol (kept at -20 ºC) for 10 min. After the ethanol was suck and washed three times with cold PBS, cells were permeabilized with 0.5% of Triton X-100 during 5 min. To reduce nonspecific background we added 1% bovine serum albumin (BSA) to the solution for 20 min. Actin filaments were stained in red with Alexa Fluor® 555 Phalloidin (Thermo Fisher Scientific) for 20 mins at 1:40 in BSA. The nucleus of both types of cells were stained with 4',6-diamidino-2-fenilindol (DAPI, ≥ 98%, Sigma-Aldrich) for 5 min, respectively, and then washed three times with cold PBS. Fluorescence microscopy of fluorescein-labelled MSN$_{PAA}$ and MSN$_{ConA}$ internalised into MC3T3-E1 and HOS cells was performed with an Evos FL Cell Imaging System (Thermo Fisher Scientific) equipped with three Led Lights Cubes (λex (nm); λem (nm)): DAPY (357/44; 447/60), GFP (470/22; 525/50), RFP (531/40; 593/40) from AMG (Advance Microscopy Group). Red cannel was used to label the cytoplasm, green for nanoparticles and blue for cell nucleus.

**2.14. Measurement of sialic acid levels**

The contents of the resulting free N-acetylneuraminic acid were measured using a sialic acid (NANA) fluorometric assay kit (BioVision Inc., Milpitas, CA, USA) according to the manufacturer's instructions, followed by detection and analyses UV-visible. This kit utilizes an enzyme coupled reaction in which free sialic acid is oxidized resulting in development of Oxi-Red probe to give absorbance (OD = 570 nm).

## 3. Results and discussion

### 3.1 Preparation and characterization of the nanosystems

Scheme 1 illustrates the different synthetic steps carried out in this work with the aim of assembling the different building blocks in to the DOX-loaded MSNs aimed at developing the final multifunctional nanodevice for antitumor therapy.

To reach this goal surfactant-containing MSN were reacted with TEPSA followed by succinic anhydride hydrolysis to attain carboxylic acid moieties on the outermost MSN surface, affording surfactant-containing $MSN_{COOH}$. Then, the acid-cleavable linker (ATU) was grafted to the external surface of such nanoparticles using the well-known carbodiimide chemistry and the surfactant was removed by solvent extraction, leading to $MSN_{ATU}$ nanosystem. The next steps consisted in loading the chemotherapeutic drug DOX inside the mesoporous cavities ($MSN_{ATU}$@DOX) and proceed to pore capping with the polymer PAA ($MSN_{PAA}$@DOX) to minimize premature cargo release before reaching the acidic environment of the endo/lysosomes of the target tumor cell. Finally, ConA was covalently linked to the resulting nanosystems, providing $MSN_{ConA}$@DOX. The proposed pH-responsive release behavior of the full nanodevice is schematically depicted as an inset in Scheme 1.

With the aim of confirming the chemical grafting of the different functional groups after each step of the synthesis, all the nanosystems were deeply characterized after surfactant removal and before DOX loading, namely, MSN, $MSN_{COOH}$, $MSN_{ATU}$, $MSN_{PAA}$ and $MSN_{ConA}$.

FTIR spectra evidence the successful functionalization stages of MSN, since vibration bands corresponding to the grafted chemical groups are observed (Fig. 1). FTIR spectrum of MSN displays vibration bands in the 490-1090 cm$^{-1}$ range typical of pure silica materials. $MSN_{COOH}$ FTIR spectrum shows an adsorption peak at 1715 cm$^{-1}$, owing to the C=O stretching vibration in carboxyl group[40]. A slight displacement of C=O stretching vibration (1698 cm$^{-1}$) and an additional absorption band located at 1636 cm$^{-1}$ characteristic of both primary and secondary amide, provided evidence of the satisfactory grafting of ATU[41]. The polymerization is evidenced by the presence of a band at 1546 cm$^{-1}$ characteristic of the asymmetric C-O stretching mode of pure PAA[42]. The incorporation of ConA onto the surface does not involve the addition of any new functional group but the variation in the proportion of those already present, that is the reason because new bands do not appear in the spectrum but the relative intensity of the signals is modified showing that reaction takes place.

The functionalization sequence was monitored by $^1H \rightarrow {}^{13}C$ CP-MAS solid-state NMR spectrum (Fig. S2, Supporting Information). Compared with pure silica MSN, where only signals corresponding to residual surfactant molecules, can be observed, $MSN_{COOH}$ showed additional resonance signals at about 15, 19, 62 and 180 ppm, which can be assigned to characteristic carbon peaks of 1,2-bidentate carboxyl groups[40]. The appearance of two signals at 80 and 110 ppm demonstrated the presence of dioxane[43] and the presence of a well-defined signal at 175 ppm indicates that ATU was covalently attached to $MSN_{COOH}$ by an amide bond. When the polymer is attached onto the surface the two distinct peaks of pure PAA, are observed: one appears at around 40 ppm and another at 181 ppm[44]. The former is the methine and methylene carbons, and the latter is the carboxyl carbon.

Table 1 summarizes some of the most relevant features of MSN, $MSN_{PAA}$ and $MSN_{ConA}$ nanosystems synthetized in this work. The differences between TGA measurements allowed determining the organic matter content incorporated in each type of nanoparticle. To determine the amount of PAA present in $MSN_{PAA}$ sample it was necessary to measure the amount of organic matter present in the different samples before polymer incorporation (results not displayed in Table 1), i.e. MSN (4.5%) $MSN_{COOH}$ (21.9%), $MSN_{ATU}$ (28.8%) and $MSN_{PAA}$ (36.2%), which allowed estimating a polymer content of *ca* 7%. Finally, *ca* 10 % was the amount of ConA existing in the final nanosystem, $MSN_{ConA}$.

Low-angle XRD patter of MSN display four resolved peaks that can be indexed as 10, 11, 20 and 21 reflections of a well-ordered 2D-hexagonal structure with p6mm plain group typical of MCM-41 (Fig. 2). Nonetheless, the intensity of such peaks XRD patters of $MSN_{PAA}$ notably decreases, even disappearing that assigned to 21 reflection, whereas those of $MSN_{ConA}$ exhibit a unique weak signal that can be assigned to 10 reflection.

These experimental results could be related to a loss of mesostructural order due to the PAA and ConA grafting processes in samples $MSN_{PAA}$ and $MSN_{ConA}$, respectively. However, it has been widely reported that it is not easy to detect alterations in the crystal structures exclusively from powder XRD[45,46]. Actually, the disappearance of the signals in XRD patterns of coated nanosystems may be also attributed to the effective filling of the mesopore channels by PAA and ConA, as it has been previously reported for MSNs coated with gelatin and decorated in the outermost surface with folic acid as targeting ligand[26]. This fact would be in good agreement with the results derived from $N_2$ adsorption porosimetry discussed below, where a decrease in the textural properties of these nanosystems occurs owing to the organic matter incorporation. Moreover, this statement is in agreement with HRTEM studies of the different samples, where the preservation of the well-ordered 2D-hexagonal structure can be clearly observed, vide infra.

As expected, the textural parameters of nanoparticles (mainly surface area, pore volume and pore diameter), derived from $N_2$ adsorption porosimetry experiments, experience a significant decrease with increasing surface decoration (Table 1). $N_2$ adsorption-desorption isotherms are of type IV corresponding to mesoporous materials (Fig. S3, Supporting Information). The suitable treatment of the experimental data evidences a dramatically decrease in the textural properties when ranging from MSN to $MSN_{PAA}$ and $MSN_{ConA}$ samples, which confirms the successful incorporation of the polymer coating and the lectin to the nanoparticles. Thus, the surface area ($S_{BET}$) decreases from 1210 m$^2$ g$^{-1}$ for MSN to 22 m$^2$ g$^{-1}$ and 15 m$^2$ g$^{-1}$ for $MSN_{PAA}$ and $MSN_{ConA}$, respectively. The initial pore volume ($V_P$) of MSN decreases from 1.41 cm$^3$ g$^{-1}$ to 0.11 cm$^3$ g$^{-1}$ and ~ 0 cm$^3$ g-1, for $MSN_{PAA}$ and $MSN_{ConA}$, respectively. Finally, the pore diameter ($D_P$) experiences a major decrease dropping from 2.4 nm for MSN to a value that could not be determined in the case of $MSN_{PAA}$ and $MSN_{ConA}$ samples. For comparison purposes, $N_2$ adsorption-desorption isotherm of $MSN_{ATU}$ sample was also registered (Fig. S2, Supporting Information), whose appropriate treatment allowed to obtain the following parameters: $S_{BET}$ = 607 m$^2$ g$^{-1}$, $V_P$ = 0.78 m$^2$ g$^{-1}$ and $D_P$ = 2.3 nm. These results clearly point to a decrease in the surface area and pore volume due to the partial blocking of the mesopore entrances ascribed to ATU moieties. However, the pore diameter is quite similar to that of MSN since the surfactant is removed after ATU anchorage to the nanoparticles. All these findings may account not only to the external polymeric covering of MSN by PAA and ConA, but also to the partial filling of the mesoporous cavities, in agreement with XRD results previously discussed.

TEM images of nanoparticles samples show a honeycomb mesoporous arrangement typical of MCM-41 (Fig. 3) Grafting of PAA and ConA does not affect the mesostructural order or MSN. The morphology of the nanoparticles is also preserved, showing spherical particles in all cases. Staining of samples with 1% PTA permitted to observe the organic matter as high contrast zones and the inorganic silica matrix as brighter areas in TEM images. The average diameter of the nanosystems estimated from the measurement of 20 nanoparticles was *ca* 160 nm, 200 and 210 nm for MSN, $MSN_{PAA}$ and $MSN_{ConA}$, respectively, with a relative error of *ca* 10%. TEM images indicate that the average thickness of the organic coating for $MSN_{PAA}$ was *ca* 4 nm (as indicated in Fig. 3, bottom left), whereas the appearance of small globular aggregates in $MSN_{ConA}$ sample (yellow arrows in Fig. 3, bottom right) accounts for the presence of the protein decorating the external polymeric layer.

To acquire information regarding the mean size and surface charge of the nanosystems, dynamic light scattering (DLS) and ζ-potential measurements were recorded (Fig. 3, top left). The mean hydrodynamic sizes determined by DLS were found to be 180 nm, 220 nm and 260 nm for MSN, $MSN_{PAA}$ and $MSN_{ConA}$, which, as expected, are slightly higher than those estimated from TEM images. This fact is explained because DLS provides the mean diameter of nanoparticles with a solvation layer[47], whereas TEM shows the size in a dry state[48].

ζ-potential measurements in water of the different nanoparticles showed notable variations in the superficial charge (Fig. S4, Supporting Information), which are consistent with the different functionalities incorporated in each step of the synthesis. The values change from -20.0 mV for MSN to -39.2 mV for $MSN_{COOH}$ because of the incorporation of –COOH groups of TEPSA onto the surface of MSN. For $MSN_{ATU}$ the ζ-potential increased to -15.0 mV what is also consistent with the incorporation of amino groups from ATU. The coating process with an acid polymer (PAA) entailed again a decrease in the ζ-potential of the so called $MSN_{PAA}$ until -54.4 mV ($MSN_{PAA}$). Finally anchoring ConA by amide linkages with the acid groups of the polymer resulted in a ζ-potential value of -34.2 mV for the entire system $MSN_{ConA}$. The different ζ-potencial values obtained account for the successfully PAA grafting and subsequent ConA anchorage in $MSN_{PAA}$ and $MSN_{ConA}$, respectively.

**3.2. In vial cargo release experiments**

$[Ru(bipy)_3]^{2+}$ was used as model molecule to evaluate the pH-responsiveness of the final nanosystem, $MSN_{ConA}$. As indicated in the experimental section, the cargo loading was carried out by dispersing 40 mg of $MSN_{ATU}$ nanoparticles into a concentrated solution of $[Ru(bipy)_3]Cl_2$ (10 mg mL$^{-1}$) and stirred at R.T. for 24 h. Then PAA capping was performed in situ via EDC-NHS chemistry. After washing and isolation of molecule-loaded $MSN_{PAA}$ samples, ConA was covalently grafted (for further details see experimental section). The estimated amount of $[Ru(bipy)_3]^{2+}$ loaded into the nanosystems was found to be 5.9% in weight. In the case of DOX, the amount of drug incorporated into the nanoparticles was 6.9% in weight. The decrease in the $V_P$ derived from $N_2$ adsorption measurements (results not shown) confirms the incorporation of the cargoes into the mesoporous channels. The pH-responsive drug delivery behavior of the final nanosystem, $MSN_{ConA}$, under physiological relevant conditions mimicking extracellular environment (pH 7.4) and those of endosomes or lysosomes (pH ≤ 5.5) $[Ru(bipy)_3]^{2+}$ was evaluated in vial. Fig. 4 shows the cargo release profiles from $MSN_{ConA}$ after soaking the nanosystem into aqueous solutions at pH 7.4 and pH 5.3 during 24 h and monitored by fluorescence excitation spectroscopy under continuous settings. Release profiles can be adjusted to first-order kinetic model by introducing an empirical non-ideality factor (δ) to give the following equation[49]:

$$Y = A(1-e^{-kt})^\delta$$

being Y the percentage of $[Ru(bipy)_3]^{2+}$ released at time t, A the maximum amount of $[Ru(bipy)_3]^{2+}$ released (in percentage), and k, the release rate constant. The values for δ are comprised between 1 for materials that obeys first-order kinetics, and 0, for materials that release the loaded cargo in the very initial time of test. The parameters of the kinetic fitting shown in Fig. 4 ($R^2 = 0.998$) indicate that, whereas the maximum amount of $[Ru(bipy)_3]^{2+}$ released is practically the totality of the loaded dye at pH 5.3, a significant amount of the cargo is retained at pH 7.4. Under acidic conditions the δ value is very close to 1, pointing to a near first order kinetics, on the contrary at the physiological pH such value is *ca* 0.6, indicating an initial burst release of the entrapped molecules. Nonetheless, at pH 5.3 the total molecule release is almost 3-fold higher than that at pH 7.4, which is associated to a k value twice greater under acidic conditions. This finding can be explained by the cleavage of the acid-labile acetal linker from ATU, which would trigger the PAA nanocoating removal and would allow the escape of the entrapped molecules in a pH-dependent controlled manner, as schematically depicted in Scheme 1 as an inset.

### 3.2. *In vitro* biological evaluation

Once $MSN_{ConA}$ sample was deeply characterized and its pH-sensitive drug release capability wad proved in vial, we proceeded to investigate its *in vitro* performance as selective drug delivery nanocarrier. To evaluate the *in vitro* biological behavior of our nanosystem and its potential application in bone cancer treatment, nanoparticles were incubated in the presence of two types of bone cells populations: MC3T3-E1 preosteoblastic cell line, and HOS human osteosarcoma cells. Herein, the grafting of ConA to the outermost surface of our nanosystems aims at playing a key role, acting as targeting ligand that permits the selective internalization of the nanocarrier by bone cancer cells and not by healthy ones. Once there, the acidic environment of the endo/lysosomes cellular compartments would trigger the release of the cytotoxic cargo.

The first step consisted in evaluating the capability of ConA-conjugated ($MSN_{ConA}$) and ConA-free ($MSN_{PAA}$) nanoparticles to be internalized by MC3T3-E1 and HOS cells. To attain this goal, both types of nanoparticles were labelled with fluorescein, as previously described in the experimental section. Cellular uptake and internalization of fluorescein-labelled $MSN_{ConA}$ and $MSN_{PAA}$ nanoparticles were investigated by flow cytometry and fluorescence microscopy in MC3T3-E1 and HOS cells in contact with (100 µg mL$^{-1}$) for 2 h (Fig. 5). After 2 h of cell culture, $MSN_{ConA}$ internalization and fluorescence intensity in HOS cells significantly increase compared to $MSN_{PAA}$, and they are always higher than those in MC3T3-E1 cells.

The cellular uptake results obtained by flow cytometry were also confirmed by fluorescence microscopy using fluorescein-labelled nanoparticles after 2 h (representative images in Fig. 6). The highest amount of internalized nanoparticles corresponds to $MSN_{ConA}$ ones after being in contact with HOS osteosarcoma cells. Notice that in this case, there is a perceptible variation in the morphology and disposition of actin filaments that make us hypothesize that those cells may be in a preapoptotic state (Fig. 6). This finding could be attributed to some cellular damage that ConA may be exerting in HOS cells (see Fig. S5, Supporting Information), as reported in the literature for various cancer cell lines[50-52]. Nonetheless, no significant cell viability decrease has been observed during cytotoxicity assays by MTS for $MSN_{ConA}$ concentrations up to 144 μg mL$^{-1}$ (Fig. S6, Supporting Information).

The next step was to investigate the ligand-receptor interactions that would be promoting preferential $MSN_{ConA}$ internalization into HOS cancer cells. In this context, in recent years, sialic acids, a type of glycans attached to glycoproteins on the cell, have been studied as disease-associated carbohydrate derivatives, because their expression provides many opportunities for the appraisal of the cell processes[53,54]. For instance, Cho et al. designed lectin-tagged fluorescent polymeric nanoparticles as potential bioimaging probes for detecting diseased cells by the union lectin-sialic acids[55]. Through cellular experiments, they successfully detected sialic acid overexpression on cancerous cells with high specificity.

Thus, to confirm the possible overexpression of sialic acid in HOS cells compared with MC3T3-E1 cells, and their interaction with ConA, we measured the N-acetylneuraminic acid levels. This acid is the most common member of sialic acid derivatives and is found widely distributed in animal tissues. We found that N-acetylneuraminic acid levels in HOS cells were significantly higher compared with the levels found in MC3T3-E1 cells: 600 ± 1 nmol μg$^{-1}$ protein in HOS cells vs. 100 ± 2 nmol μg$^{-1}$ protein in MC3T3-E1 cells. These results suggest that the presence of ConA in the surface of the nanoparticles allows the binding with sialic acid and improving the selective internalization of $MSN_{ConA}$ in HOS cells where the sialic acid levels were 6-fold increased.

Once demonstrated the capability of $MSN_{ConA}$ nanoparticles to be preferentially internalized by HOS cells than by MC3T3-E1 cells, the next goal was to investigate and compare the in vitro performance of DOX-loaded lectin-conjugated ($MSN_{ConA}$@DOX) and lectin-free ($MSN_{PAA}$@DOX) nanosystems in cultures of both cell types at different nanoparticles concentrations. However, evaluating the killing capability of free DOX towards both cell types is necessary to compare the efficiency and selectivity of the here developed nanotransporter. Thus, different amounts of free DOX (0, 2.5, 5, 10 and 20 µg mL$^{-1}$) were used, and the results indicate that cell death induced by DOX is very high and shows a linear dose-dependent concentration both in preosteoblastic MC3T3-E1 and osteosarcoma HOS cells (Fig. S7, Supporting Information). This accounts for the high and unselective cytotoxicity of this drug.

Then, we performed and *in vitro* cytotoxicity study (by a MTS assay) with DOX concentrations (2.5 and 10 µg mL$^{-1}$) incorporated in MSN$_{ConA}$ and MSN$_{PAA}$ nanoparticles in the presence of MC3T3-E1 or HOS cells at 48 h (Fig. 7). Comparing the toxicity of MSN$_{ConA}$@DOX vs. that of MSN$_{PAA}$@DOX, it is observed that MSN$_{ConA}$@DOX causes increased cell death than MSN$_{PAA}$@DOX, and that this increase is much more significant in HOS cells that in MC3T3-E1. Actually, it should be highlighted that this effect is much more obvious at very low DOX concentrations (2.5 µg mL$^{-1}$). Thus, after 48 h of assay MC3T3-E1 cells does not experience any cytotoxic effect after being incubated with 2.5 µg mL$^{-1}$ of DOX loaded in both nanosystems. Oppositely, HOS cells viability decreases to *ca* 40% after exposition to 2.5 µg mL$^{-1}$ of drug loaded into MSN$_{PAA}$. This value is quite similar to that for HOS cells after exposure to free DOX (Fig. S7, Supporting Information). Nonetheless, the antitumor effect of MSN$_{ConA}$ towards HOS cells increases 8-fold compared to MSN$_{PAA}$ (Fig. S7, Supporting Information).

## 4. Conclusions

In this work we have developed a multifunctional nanodevice featuring selectivity towards human osteosarcoma cells and pH-responsive antitumor drug delivery capability. The synergistic combination of both properties into a unique nanosystem produces an amplification of the antitumor efficacy.

This innovative nanodevice is based in DOX-loaded MSNs nanoplatforms where different building blocks are assembled: i) a PAA polymeric shell, anchored *via* an acid cleavable linker, to prevent premature cargo release and provide the nanosystem of pH-responsive capability; ii) a targeting ligand consisting in the lectin ConA grafted to PAA, to increase the selectivity towards cancer cells whereas significantly preserving the viability of healthy cells.

*In vitro* assays reveal that the internalization degree of lectin-conjugated nanosystems into human osteosarcoma cells is 2 times higher than in human preosteoblastic cells. Moreover, only very small DOX concentrations (2.5 µg mL$^{-1}$) are needed to attain *ca* 100% antitumor efficacy against osteosarcoma cells compared to healthy bone cells, whose viability is preserved. Moreover, the antitumor effect is increased up to 8-fold compared to that caused by the free drug.

These outcomes prove that the synergistic assembly of different building blocks into a unique nanoplatform increases antitumor effectiveness and decreases toxicity towards healthy cells, which constitutes a new paradigm in targeted bone cancer therapy.

**Apendix A. Supplementary data**

Supplementary data associated with this artice can be found in the online version.

**Acknowledgements**

MVR acknowledges funding from the European Research Council (Advanced Grant VERDI; ERC-2015-AdG Proposal No. 694160). The authors also thank to Spanish and Ministerio de Economía y Competitividad (MINECO) (Project MAT2015-64831-R). The XRD measurements, $^1$H → $^{13}$C solid-state NMR spectra and internalization studies were performed at C.A.I. Difracción de Rayos X, C.A.I. Resonancia Magnética Nuclear and C.A.I. Citometría de Flujo from UCM (Spain), respectively. TEM studies were performed at ICTS National Centre for Electron Microscopy (Spain).

**STATEMENT OF SIGNIFICANCE**

The development of highly selective and efficient tumor-targeted smart drug delivery nanodevices remains a great challenge in nanomedicine. This work reports the design and optimization of a multifunctional nanosystem based on mesoporous silica nanoparticles (MSNs) featuring selectivity towards human osteosarcoma cells and pH-responsive antitumor drug delivery capability. The novelty and originality of this manuscript relies on proving that the synergistic assembly of different building blocks into a unique nanoplatform increases antitumor effectiveness and decreases toxicity towards healthy cells, which constitutes a new paradigm in targeted bone cancer therapy.

Prof. Dr. María Vallet-Regí,
Departamento de Química Inorgánica y Bioinorgánica
Facultad de Farmacia
Universidad Complutense
28040, Madrid, Spain
Phones: +34-913941861/3941843; Fax: +34-91-3941786
e-mail: vallet@ucm.es
Web : https://www.ucm.es/valletregigroup

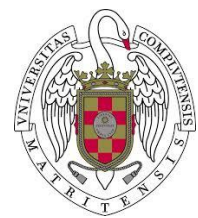

Madrid, July 28th 2017

Dear Editor,

Please find herewith the manuscript entitled: "**New Paradigm in Targeted Bone Cancer Therapy**" submitted to be considered for publication as full paper in **Acta Biomaterialia**.

Corresponding author: María Vallet-Regí

Co-corresponding author: Montserrat Colilla

Type of manuscript: Full Length Article

Postal address: Departamento de Química Inorgánica y Bioinorgánica, Facultad de Farmacia. Universidad Complutense de Madrid, Plaza Ramón y Cajal s/n, 28040 Madrid, Spain.

E-mail addresses:

    Corresponding author: vallet@ucm.es

    Co-corresponding author: mcolilla@ucm.es

Phone: +34 91 394 1861. Fax: +34 91 394 1786

Co-authors: Marina Martínez-Carmona and Daniel Lozano

The major constraint of conventional chemotherapy for cancer treatment is the lack of specificity of cytotoxic drugs. This constitutes a serious threat to healthy tissues, resulting in a significant decrease of the efficacy and provoking the apparition of side-effects in the patient associated to systemic toxicity. The emergence of nanotechnology has transformed this scenario owing to the development of nanocarriers for therapeutic agents.

Among nanocarriers, mesoporous silica nanoparticles (MSN) are receiving growing interest since they exhibit unique properties. They can host different therapeutic payloads and diverse organic functionalities can be attached to their surface, such as stimuli-responsive gatekeepers, to avoid premature cargo leakage until the presence of a given stimulus triggers drug release, or targeting ligands to guide MSNs towards diseased tissues.

Although some targeting ligands, such as transferrin, certain peptides, hyaluronic acid, or folic acid, have been incorporated into stimuli-responsive MSNs, the development of highly selective and efficient tumor-targeted smart drug delivery nanodevices remains a tremendous challenge.

In this work we have developed a multifunctional nanodevice featuring selectivity towards human osteosarcoma cells and pH-responsive antitumor drug delivery capability. The synergistic combination of both properties into a unique nanosystem produces an amplification of the antitumor efficacy.

This innovative nanodevice is based in doxorubicin (DOX)-loaded MSNs nanoplatforms where different building blocks are assembled: *i)* a PAA polymeric shell, anchored via an acid cleavable linker, to prevent premature cargo release and provide the nanosystem of pH-responsive capability; *ii)* a targeting ligand consisting in the lectin concanavalinA (ConA) grafted to PAA, to increase the selectivity towards cancer cells whereas significantly preserving the viability of healthy cells.

*In vitro* assays reveal that the internalization degree of lectin-conjugated nanosystems into human osteosarcoma cells is 2 times higher than in human preosteoblastic cells. Moreover, only very small DOX concentrations are needed to attain *ca.* 100% antitumor efficacy against osteosarcoma cells compared to healthy bone cells, whose viability is preserved. Moreover, the antitumor effect is increased up to 8-fold compared to that caused by the free drug.

These outcomes prove that the synergistic assembly of different building blocks into a unique nanoplatform increases antitumor effectiveness and decreases toxicity towards healthy cells, which constitutes a new paradigm in targeted bone cancer therapy.

**The Authors believe that this manuscript meets the scope of Acta Biomaterialia, and really think that the results derived from the paper are of enough quality, novelty and scientific impact to be worth of publication in this prestigious journal.**

The manuscript contains 1 scheme, 7 figures and 1 table. In the manuscript there should be enough data fulfilling the quality standards of *Acta Biomaterialia*. Supporting Information (containing 7 figures), 1 table and a synopsis graph are enclosed with the article.

The manuscript, or its contents in any other form, is not under consideration for publication and has not been published elsewhere in any medium including electronic journals and computer databases of a public nature.

Thank you for your kind consideration of our work.

Best regards

The authors

As possible referees we suggest:

**1) Prof. Dr. Jesús Martínez de la Fuente**

Departamento 4: Materiales multifuncionales y biomateriales, Instituto de Ciencia de Materiales de Aragón, ICMA. Zaragoza, Spain.

jmfuente@unizar.es

**2) Aldo Boccaccini**

Department of Materials Science and Engineering, Institute of Biomaterials, University of Erlangen-Nuremberg, 91058 Erlangen, Germany

E-mail: aldo.boccaccini@fau.de

**3) Prof. Valentina Cauda**

Politecnico di Torino, Department of Applied Science and Technology, Torino, Italy

E-mail: valentina.cauda@polito.it

**Graphical Abstract**

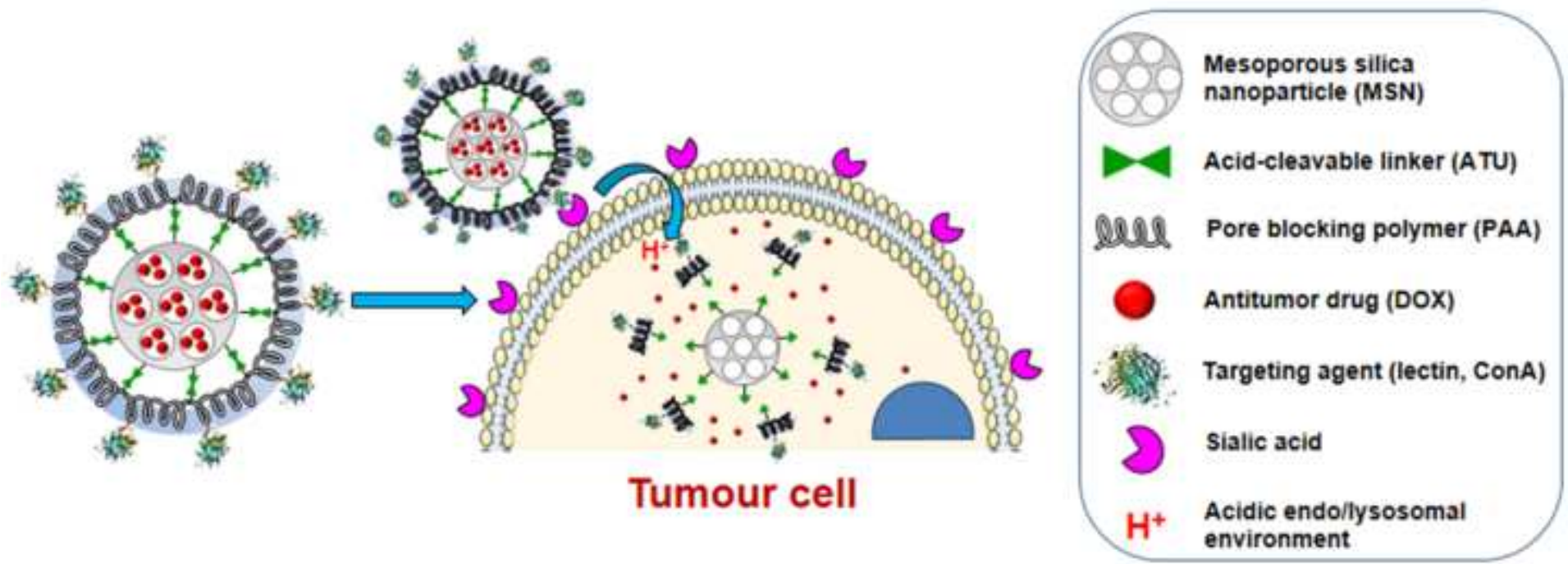

**Scheme 1**
[Click here to download high resolution image](#)

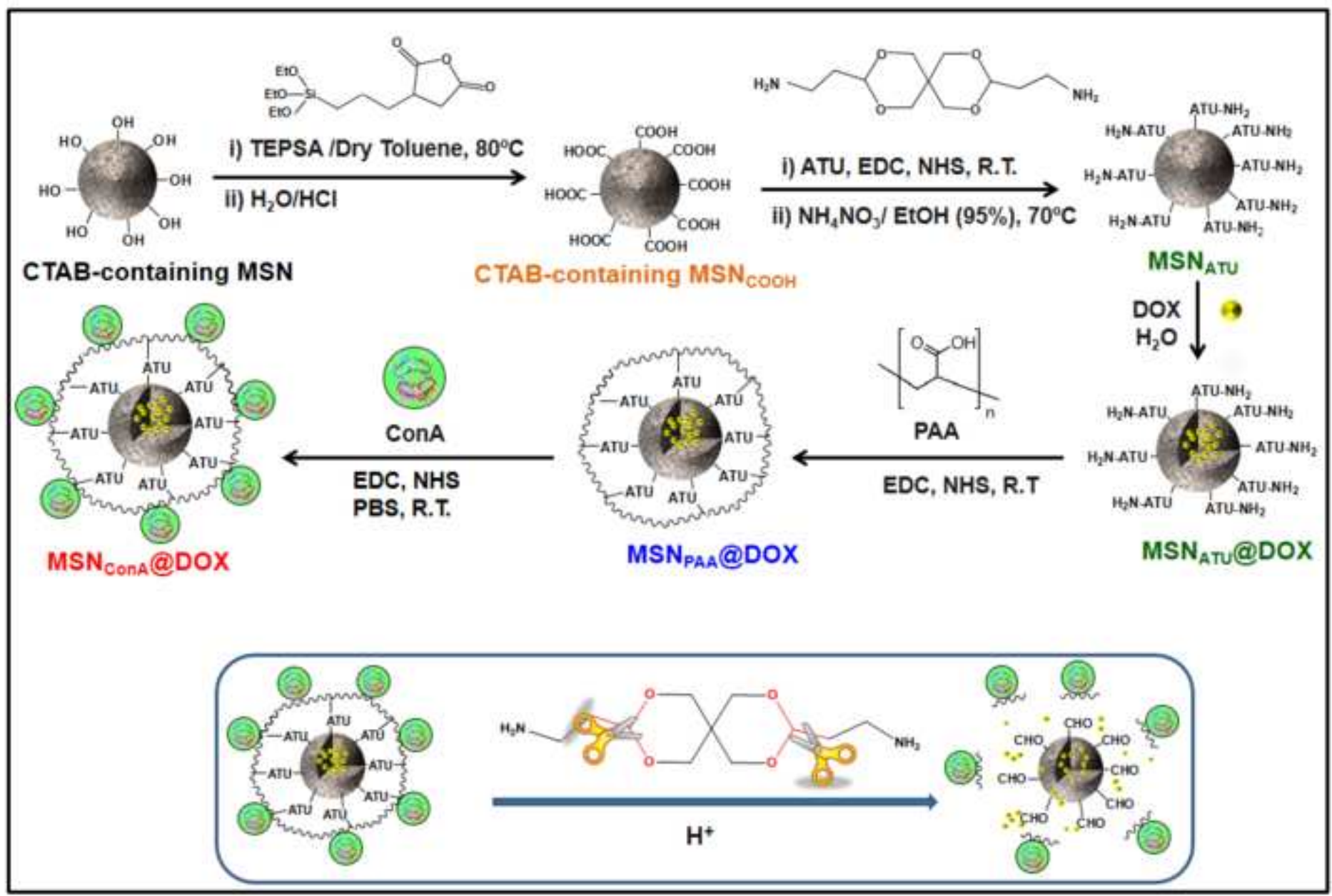

**Figure 1**
**Click here to download high resolution image**

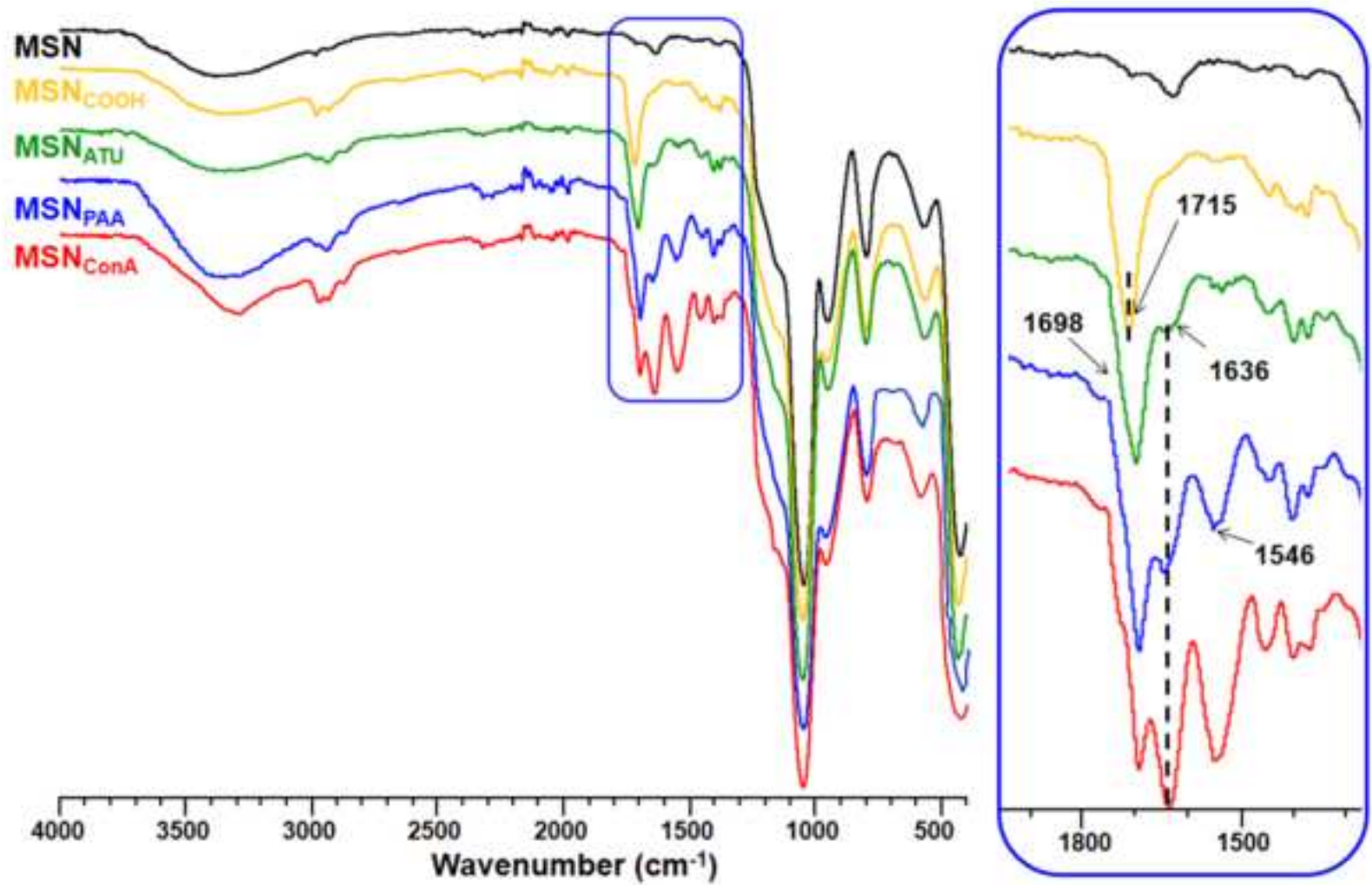

**Figure 2**
**Click here to download high resolution image**

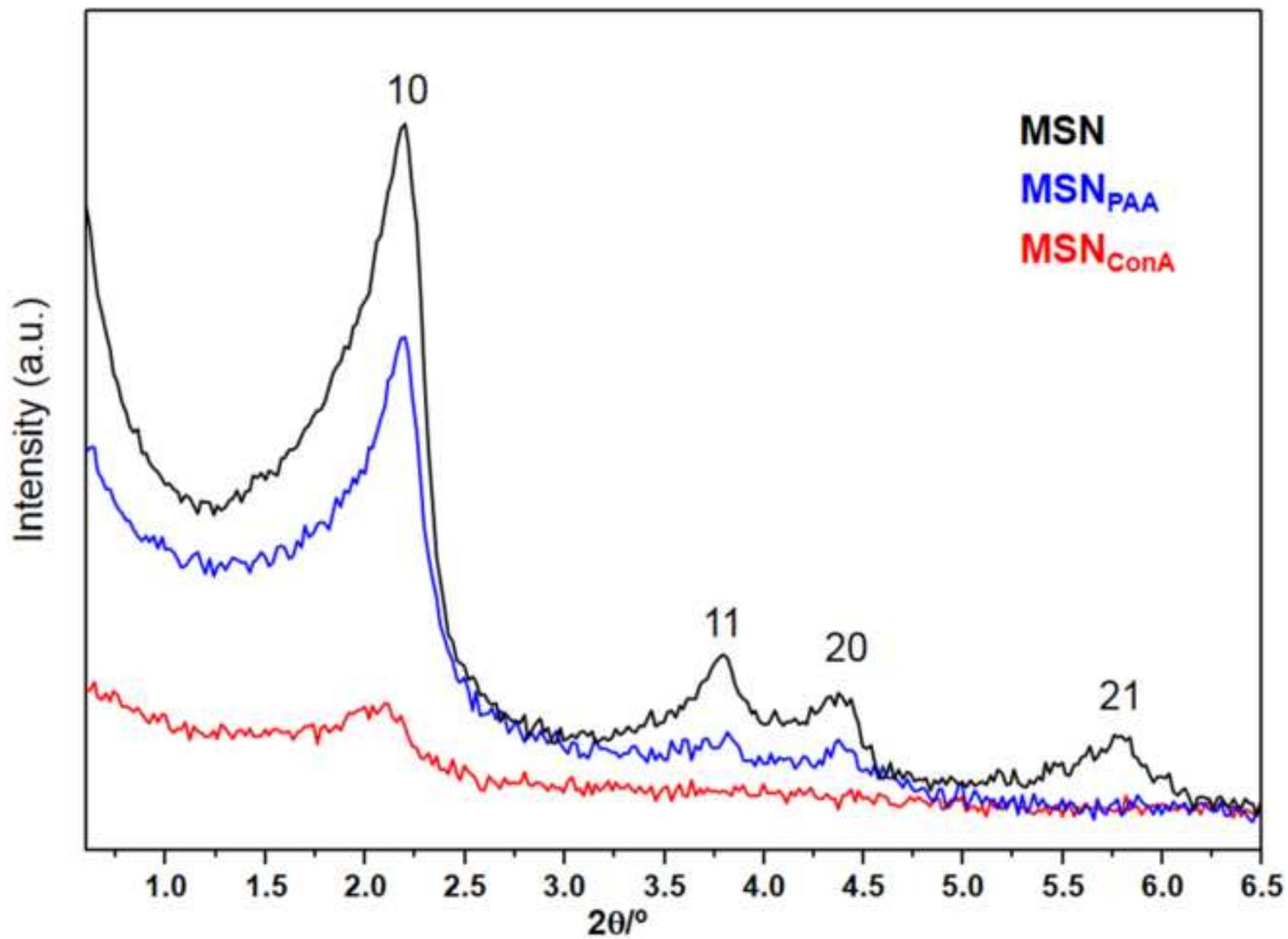

**Figure 3**
Click here to download high resolution image

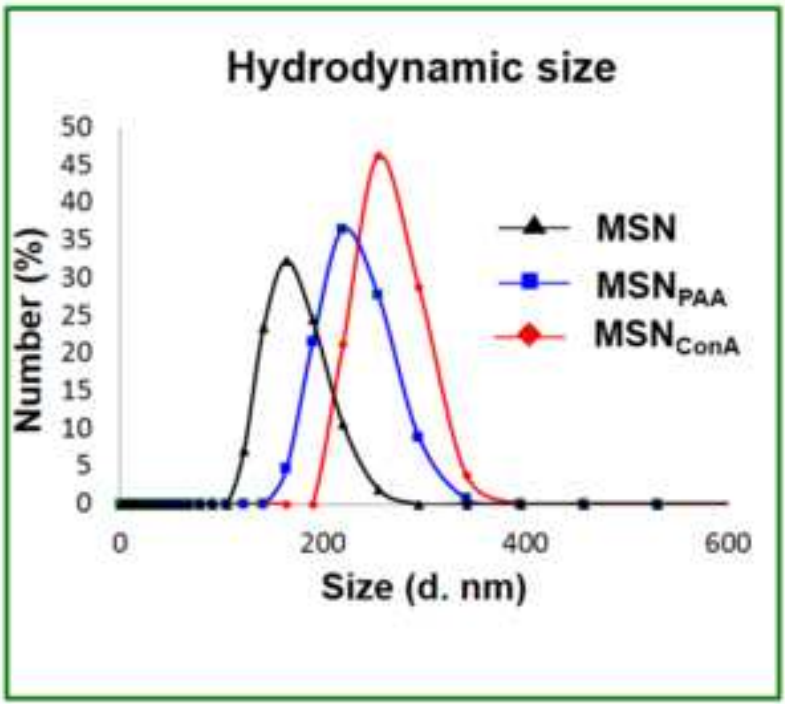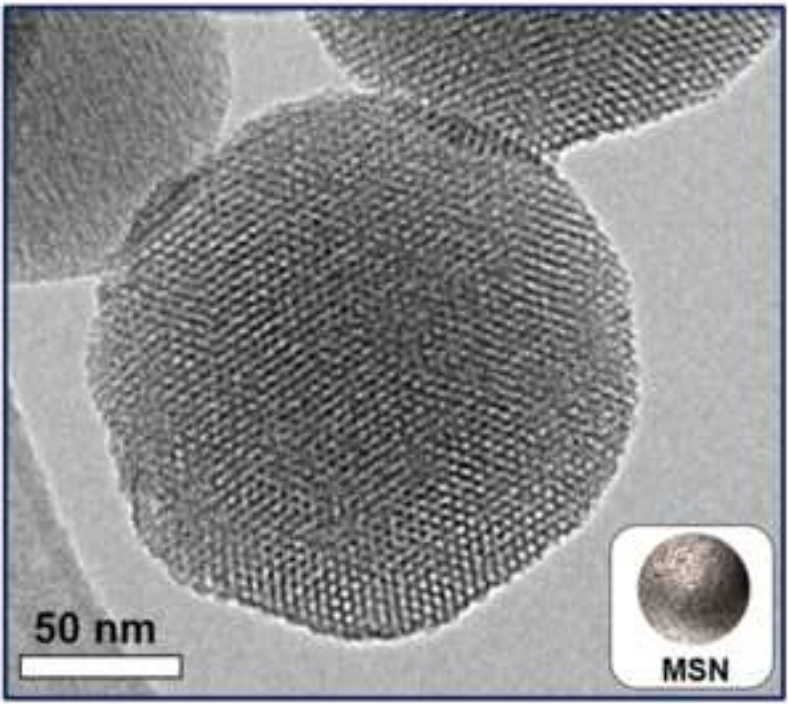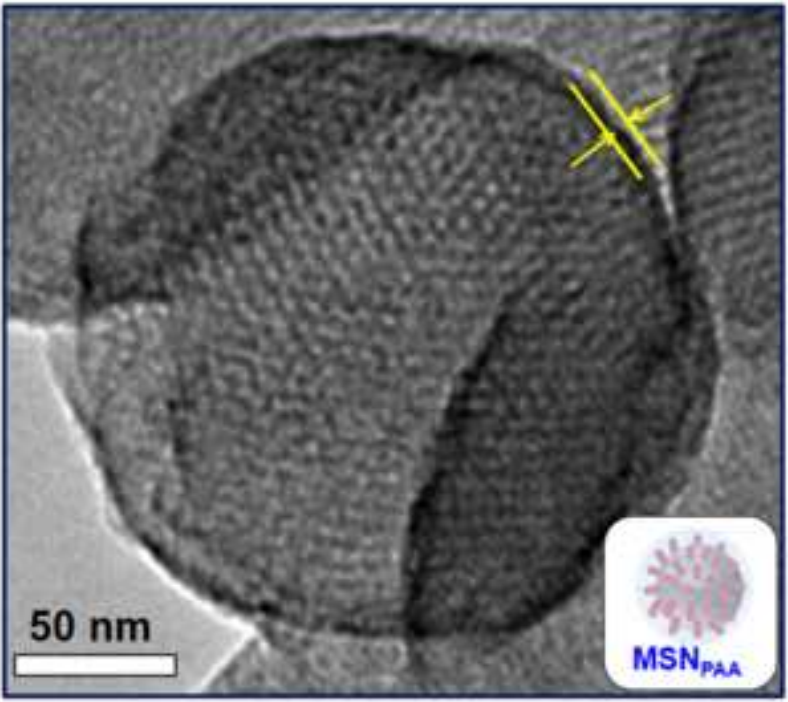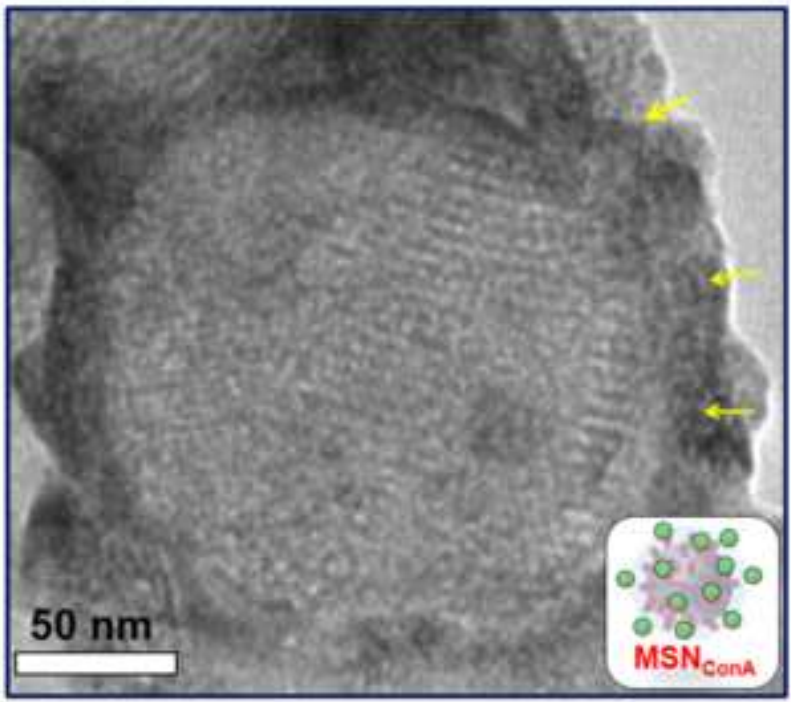

**Figure 4**
[Click here to download high resolution image](#)

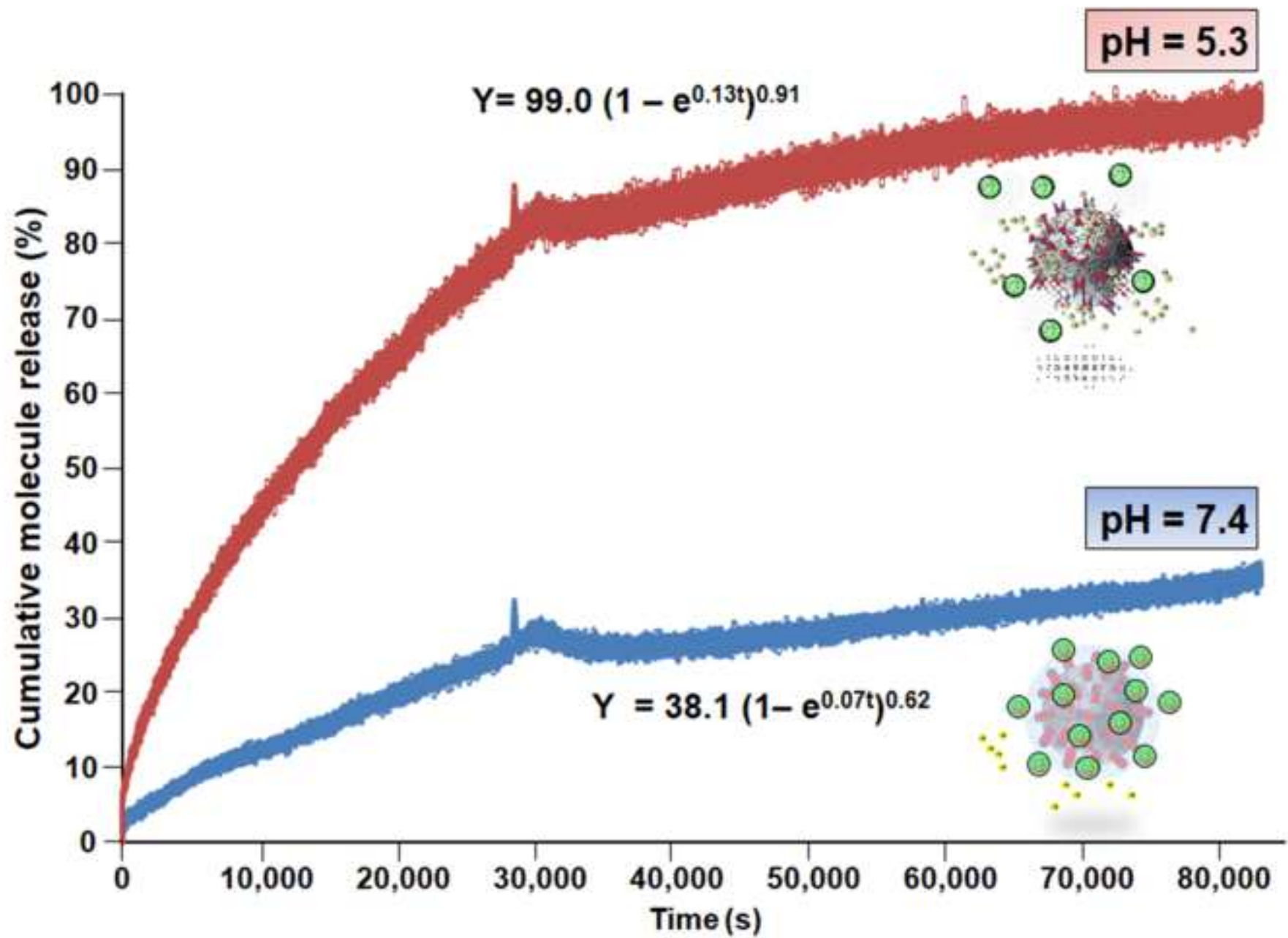

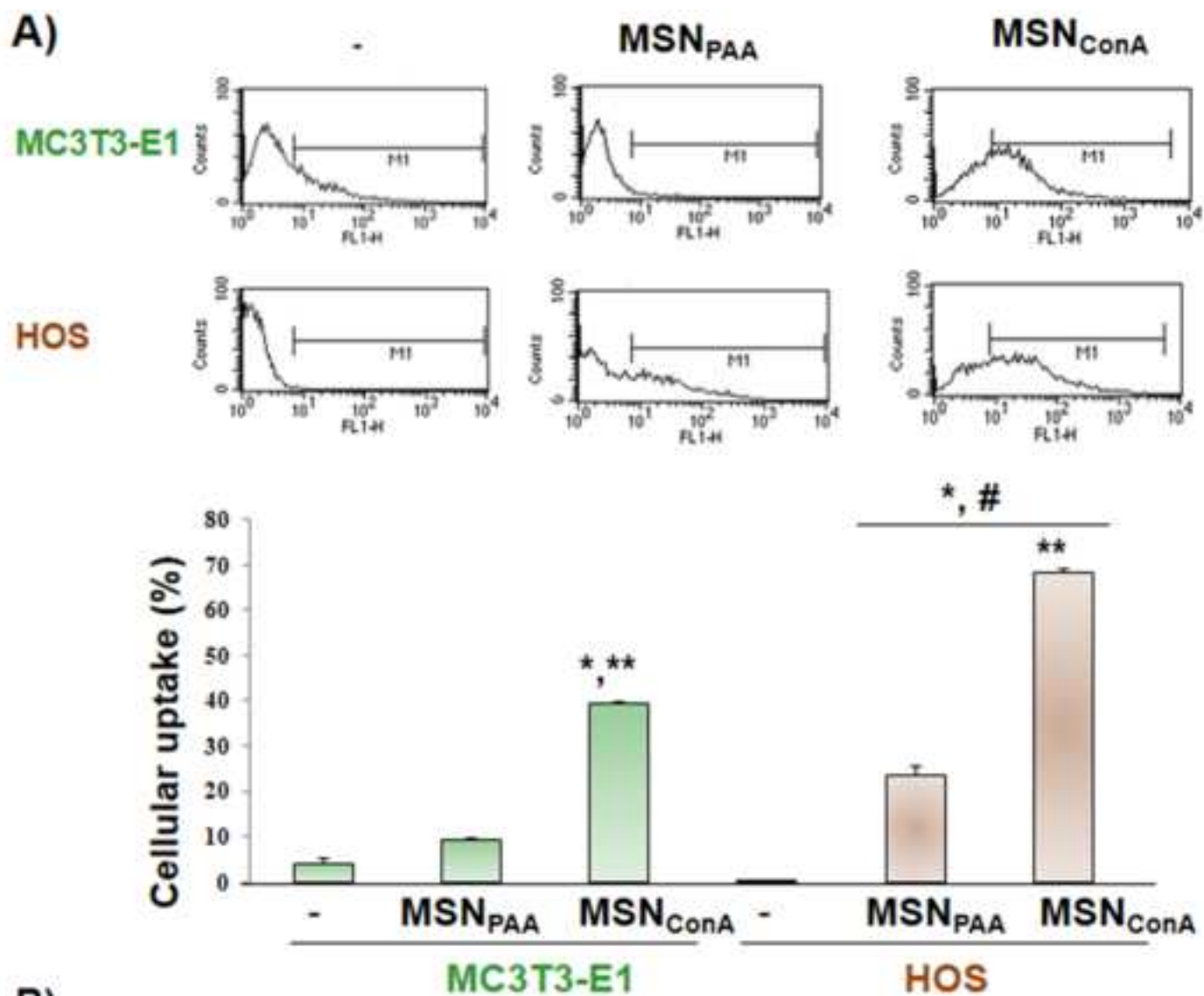
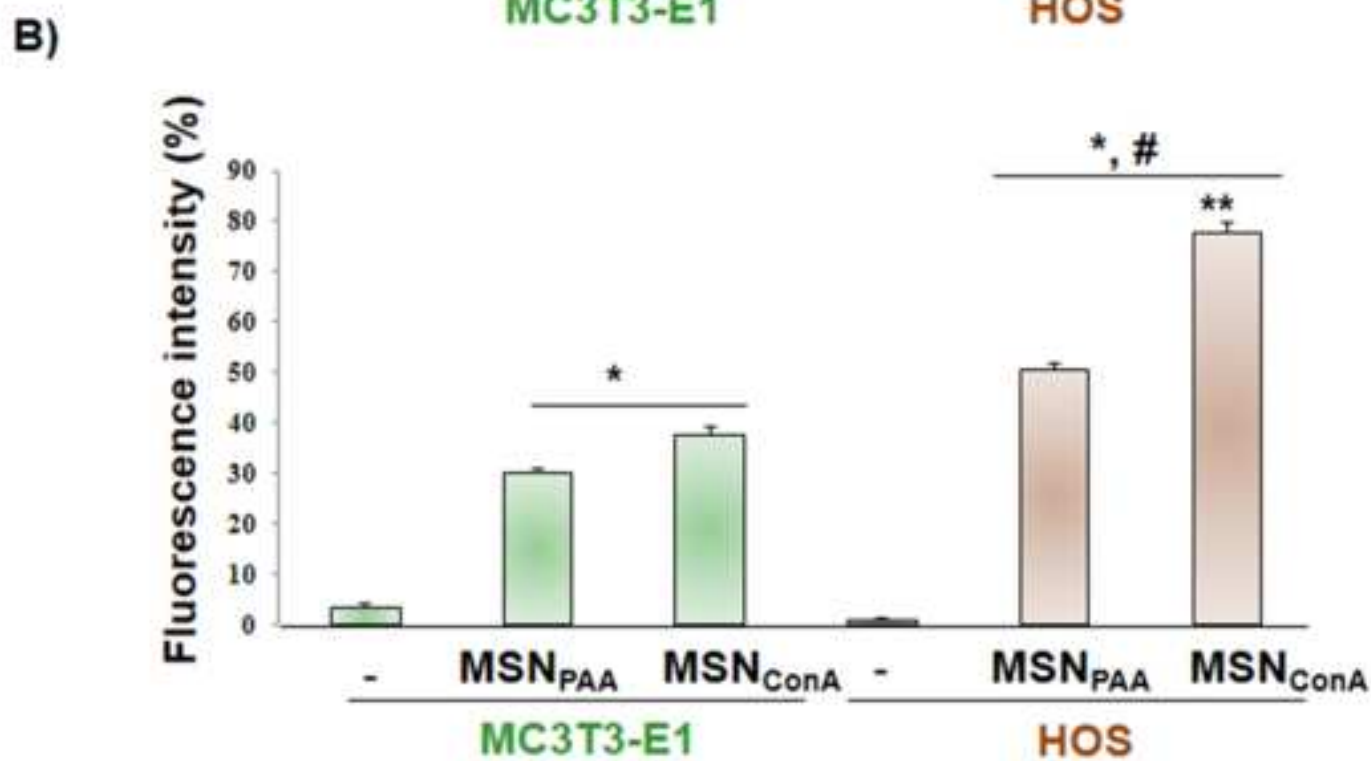

**Figure 6**
**Click here to download high resolution image**

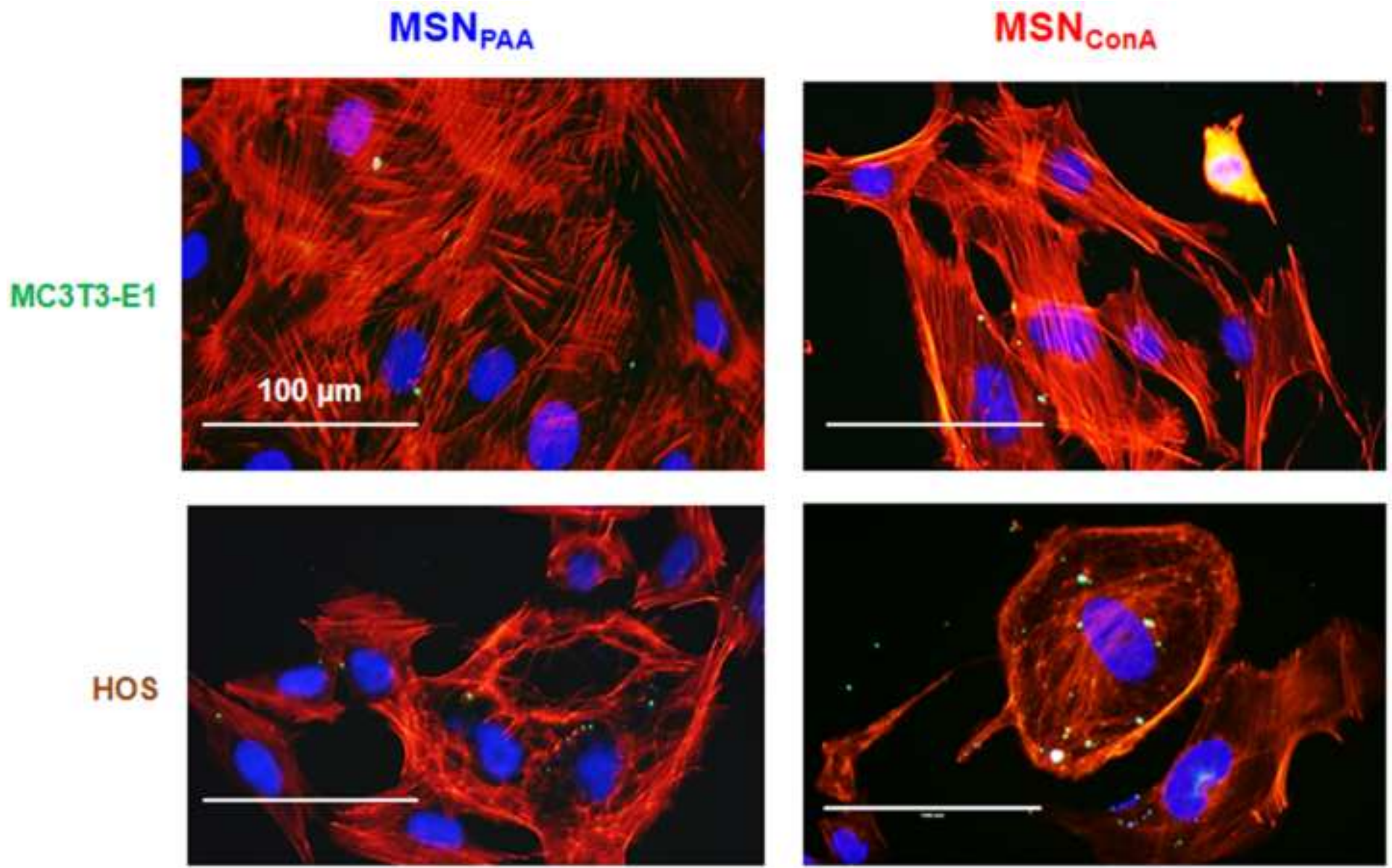

**Figure 7**
**Click here to download high resolution image**

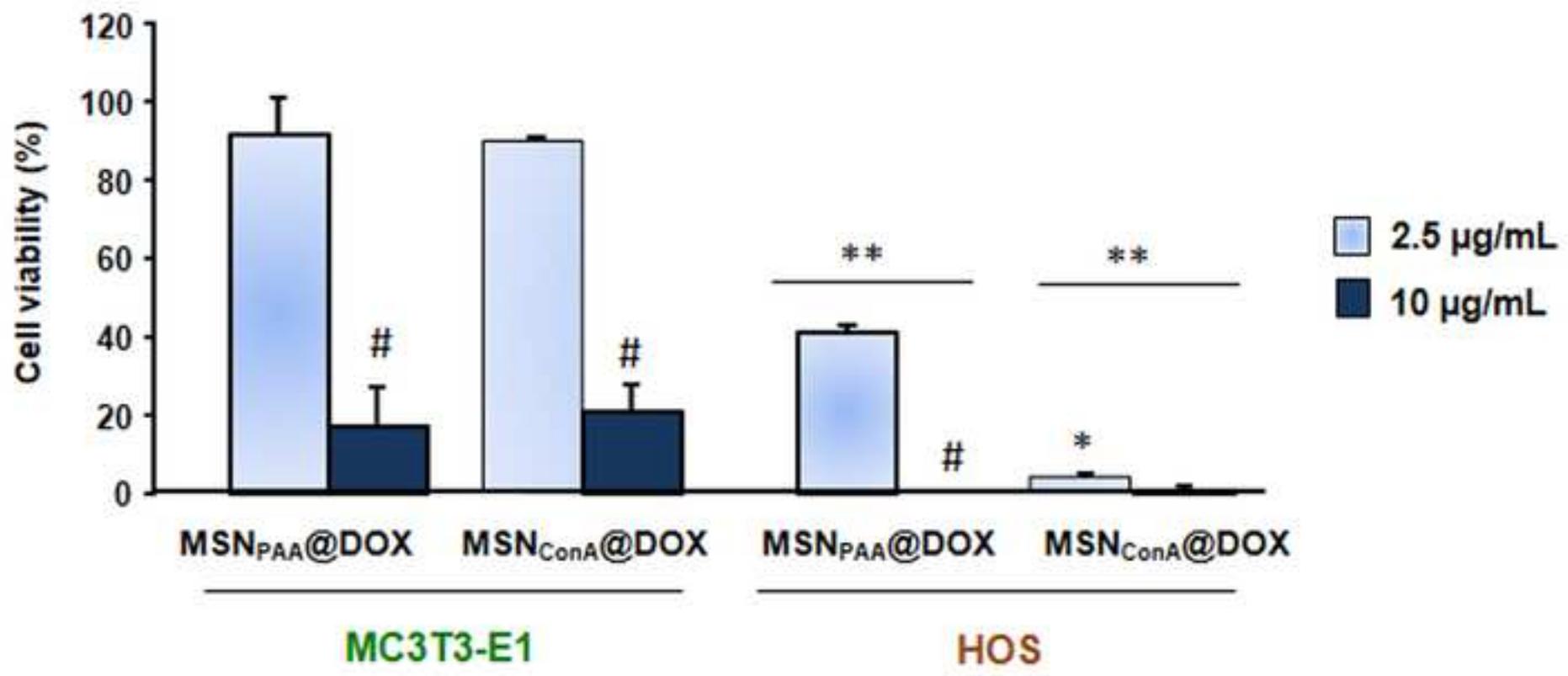

**Table 1.** Main properties of nanosystems synthetized in this work.

| Material | MSN | MSN$_{PAA}$ | MSN$_{ConA}$ |
|---|---|---|---|
| Organic matter (TGA) (%) | 4.5 ± 0.1 | 36.2 ± 0.7 | 46.5 ± 0.9 |
| $S_{BET}$ (m$^2$ g$^{-1}$) | 1210 ± 35 | 22 ± 1 | 12 ± 1 |
| $V_P$ (cm$^3$ g$^{-1}$) | 1.41 ± 0.03 | 0.11 ± 0.01 | ~ 0 |
| $D_P$ (nm) | 2.4 ± 0.1 | - | - |
| ζ-potential (mV) | -20.0 ± 1.0 | -54.4 ± 4.0 | -34.2 ± 0.5 |
| Mean size (DLS) (nm) | 180 ± 5 | 220 ± 8 | 260 ± 10 |

**Supporting Information**

[Click here to download Supplementary Material: Supporting Information Acta Biomaterialia-28-07-2017.docx](#)